\documentclass[aps,prb,twocolumn,showpacs,showkeys,superscriptaddress]{revtex4-1}

\usepackage{graphics}
\usepackage{color}
\usepackage{amsmath}
\usepackage{bm}

\input{MyMc}

\newcommand{\dX}{\ensuremath{\Delta X}\xspace}
\newcommand{\dk}{\ensuremath{\delta\kappa}\xspace}
\newcommand{\beak}{\ensuremath{{\diamondsuit}}}

\begin{document}

\title{%
  Effective model for a short Josephson junction with a phase discontinuity.
}

\author{E. Goldobin}
\email{gold@uni-tuebingen.de}
\affiliation{%
  Physikalisches Institut and Center for Collective Quantum Phenomena in LISA$^+$,
  Universit\"at T\"ubingen, Auf der Morgenstelle 14, D-72076 T\"ubingen, Germany
}

\author{S. Mironov}
\email{sermironov@rambler.ru}
\altaffiliation[Present address:]{
  Moscow Institute of Physics and Technology, 147700,
  Dolgoprudny, Russia
}
\affiliation{%
  LOMA, UMR-CNRS 5798,  Universit\'e Bordeaux,
  351, cours de la Liberation, F-33405 Talence Cedex,
  FRANCE
}
\author{A. Buzdin}
\email{a.bouzdine@loma.u-bordeaux1.fr}
\affiliation{%
  LOMA, UMR-CNRS 5798,  Universit\'e Bordeaux,
  351, cours de la Liberation, F-33405 Talence Cedex,
  FRANCE
}

\author{R.G. Mints}
\email{mints@post.tau.ac.il}
\affiliation{%
  The Raymond and Beverly Sackler School of Physics and Astronomy,
  Tel Aviv University, Tel Aviv 69978, Israel
}

\author{D. Koelle}
\email{koelle@uni-tuebingen.de}
\author{R. Kleiner}
\email{kleiner@uni-tuebingen.de}
\affiliation{%
  Physikalisches Institut and Center for Collective Quantum Phenomena in LISA$^+$,
  Universit\"at T\"ubingen, Auf der Morgenstelle 14, D-72076 T\"ubingen, Germany
}

\date{%
  \today
}

\begin{abstract}

  We consider a short Josephson junction with a phase discontinuity $\kappa$ created, \eg, by a pair of tiny current injectors, at some point $x_0$ along the length of the junction. We derive the effective current-phase relation (CPR) for the system as a whole, \ie, reduce it to an effective point-like junction. From the effective CPR we obtain the ground state of the system and predict the dependence of its critical current on $\kappa$. We show that in a large range of $\kappa$ values the effective junction behaves as a $\varphi_0$ Josephson junction, \ie, has a unique ground state phase $\varphi_0$ within each $2\pi$ interval. For $\kappa\approx\pi$ and $x_0$ near the middle of the junction one obtains a $\varphi_0\pm\varphi$ junction, \ie, the Josephson junction with degenerate ground state phase $\varphi_0\pm\varphi$ within each $2\pi$ interval. Further, in view of possible escape experiments especially in the quantum domain, we investigate the scaling of the energy barrier and eigenfrequency close to the critical currents and predict the behavior of the escape histogram width $\sigma(\kappa)$ in the regime of the macroscopic quantum tunneling.

\end{abstract}

\pacs{
  74.50.+r,   
  85.25.Cp    
}

\keywords{current-phase relation}

\maketitle

\section{Introduction}

Recently a lot of attention is attracted to Josephson junctions (JJs) with an unconventional current-phase relation (CPR)\cite{Golubov:2004:CPR,Buzdin:2005:Review:SF}. In particular, $\varphi_0$ JJs\cite{Krive:2005:S-LL-S,Buzdin:2008:varphi0,Reynoso:2008:SpinPol-QPC:JosEff,Alidoust:2013:Geom-varphi0,Goldobin:2015:0-pi-SQUID=varphi-JJ,Mironov:2015:S-NW2-S} and $\varphi$ JJs\cite{Mints:1998:SelfGenFlux@AltJc,Mints:2002:SplinteredVortices@GB,Buzdin:2003:phi-LJJ,Goldobin:CPR:2ndHarm,Sickinger:2012:varphiExp,Bakurskiy:2013:S-NF-S:varphi-JJ,Heim:2013:varphi-JJ/S-FN-S} and their combinations\cite{Goldobin:2015:0-pi-SQUID=varphi-JJ}, proposed and/or demonstrated recently, show non-trivial physics\cite{Goldobin:2013:RetrapButfly} and have potential for applications in the classical\cite{Ortlepp:2006:RSFQ-0-pi,Khabipov:2010:RSFQ:pi-flip-flop,Goldobin:2013:varphi-bit} and the quantum domains\cite{Feofanov:2010:SFS:pi-qubit} similar to $\pi$ JJs. Here, $\varphi_0$ JJs are defined as JJs having a unique ground state phase (single Josephson energy minimum situated at) $\varphi_0\neq0$ within each $2\pi$ phase interval, while $\varphi$ JJs (sometimes denoted also $\pm\varphi$ JJ) have a doubly degenerate ground state phase (double-well Josephson energy with minima at) $\pm\varphi$ within each $2\pi$ interval.

Currently, the classical properties of $\varphi$ JJs made of a short 0-$\pi$ JJ are understood rather well\cite{Goldobin:CPR:2ndHarm,Sickinger:2012:varphiExp,Goldobin:2013:varphi-bit}. For example, $\varphi$ JJs have two critical currents $I_{c-}$ and $I_{c+}$ corresponding to the escape of the phase from $-\varphi$ and $+\varphi$ wells. In our group we are starting investigation of quantum properties of such JJs. The first step in this direction could be an observation of the macroscopic quantum tunneling (MQT) of the phase\cite{Voss:1981:JJ.MQT,Bauch:2005:YBCO-45GBJJ:MQT,Li:2007:BSCCO:MQT} out of \emph{both} $-\varphi$ and $+\varphi$ wells of the Josephson energy profile. For this purpose, one, usually, measures the phase escape statistics, by sweeping the bias current at a constant rate and measuring the exact value of the switching current many times. Assuming that in $\varphi$ JJ at low temperatures (low damping) the initial state ($-\varphi$ or $+\varphi$) is random\cite{Goldobin:2013:RetrapButfly}, the switching current histogram should have two peaks, each of them just below corresponding critical current  $I_{c\pm}$. The widths $\sigma(T)$ of each histogram peak usually (when the damping is small) decreases with decreasing temperature $T$. However, $\sigma(T)$ is expected to saturate at some value $\sigma_\mathrm{min}$ for temperature below some $T^*$. Such behavior is usually interpreted as a transition from the regime of the thermal activation of the phase over the barrier to the regime of the MQT of the phase through the barrier. However, it is necessary to show that the observed $\sigma_\mathrm{min}$ is not related to the (noise in the) experimental setup and other trivial reasons. Usually, in such experiments one introduces some extra tuning parameter, \eg, a magnetic field, which allows to demonstrate that the setup is able to measure the histograms that are more narrow than $\sigma_\mathrm{min}$. Simultaneously, for the MQT experiment with $\varphi$ JJ, it would be advantageous to have a tuning parameter, which provides a continuous transition between $\varphi$ (or $\varphi_0$) JJ and a conventional 0 JJ, whose physics is well studied.

For this purpose, we propose to use a short 1D conventional 0 JJ, equipped with a pair of tiny current injectors. By sending a current $I_\mathrm{inj}$ from one injector to the other, we can create a $\kappa$ discontinuity of the Josephson phase at some point $x_0$ along the JJ, where injectors are attached\cite{Goldobin:2004:Art-0-pi,Gaber:2005:NonIdealInj2,Ustinov:2002:ALJJ:InsFluxon,Malomed:2004:ALJJ:Ic(Iinj)}. If $\kappa=\pi$, the system is similar to a superconductor-insulator-ferromagnet-superconductor (SIFS) 0-$\pi$ JJ, which becomes a $\varphi$ JJ, if parameters are chosen correctly\cite{Goldobin:2011:0-pi:H-tunable-CPR,Lipman:2014:varphiEx}. However, since $\kappa\propto I_\mathrm{inj}$ is adjustable, one can \textit{in situ} tune the junction from a 0 JJ to a $\varphi$ JJ and also study all the states in between. Tuning $\kappa$ one can also affect the widths $\sigma_\mathrm{min}$ of the hystograms.

The aim of this work is to develop a theoretical model for a short JJ with $\kappa$ discontinuity of the phase and to predict or interpret the results of MQT experiment such as the one outlined above. Namely, we derive an effective (averaged) CPR for a short JJ with a phase discontinuity $\kappa$ and obtain experimentally relevant quantities, such as the critical current or the escape histogram width as a function of $\kappa$.

The paper is organized as follows. In Sec.~\ref{Sec:Model} we introduce the model and present the averaged CPR and the averaged Josephson energy derived in details in appendix \ref{Sec:Derivation}. In Sec.~\ref{Sec:Results} we obtain several experimentally relevant dependences such as the ground state phase, critical current and escape-related characteristics as functions of $\kappa$. Sec.~\ref{Sec:Conclusions} concludes the work.

\section{Model}
\label{Sec:Model}

We consider a short JJ of length $2w$ ($w<1$ in units of Josephson length $\lambda_J$). The Josephson phase $\phi(x)$ changes along the $x$ coordinate ($-w<x<+w$). At $x=x_0$ ($-w<x_0<+w$) there is a $\kappa$ discontinuity of the Josephson phase, created, \eg, by a pair of tiny (in theory infinitesimal) current injectors\cite{Gaber:2005:NonIdealInj2}. The junction is biased by a uniform current density $\gamma$ (given in the units of the critical current density). Our aim is to derive an effective (averaged over the JJ length) current-phase relation for this system, \ie, $\gamma(\psi)$, where
\begin{equation}
  \psi\equiv\av{\phi(x)} = \frac{1}{2w}\int_{-w}^{+w} \phi(x) \,dx
  , \label{Eq:psi.def}
\end{equation}
is the average phase across the JJ. It is $\psi$ that is actually measured, if one considers the system described above as a black box with two electrodes.

In appendix \ref{Sec:Derivation} we derive the averaged CPR of the system under question by using the perturbation theory up to the second order in $w$, treating  $w$ (the half-length of the JJ) as a small parameter. It is convenient to write the resulting averaged CPR as function of the phase $\theta$, which is related to the average phase $\psi$ across the JJ as
\begin{equation}
  \theta = \psi + \frac\kappa2 X_0
  , \label{Eq:theta.def}
\end{equation}
where $X_0=x_0/w$. The averaged CPR can be written as
\begin{equation}
  \gamma(\theta) = \gamma_0(\theta) + w^2 \gamma_2(\theta) + O(w^4)
  , \label{Eq:gamma}
\end{equation}
where
\begin{equation}
  \gamma_0(\theta) = \cos\left( \frac\kappa2 \right) \sin(\theta) - X_0 \sin\left( \frac\kappa2 \right) \cos(\theta)
  , \label{Eq:gamma0}
\end{equation}
is the 0-th order result of the perturbation theory, the first order gives no correction, and
\begin{equation}
  \gamma_2(\theta) = \frac{Q}{w^2} \sin^2\left( \frac\kappa2 \right) \sin(2\theta)
  , \label{Eq:gamma2}
\end{equation}
is the second order correction in the framework of the perturbation theory, and $Q=(w^2/6)(1-X_0^2)^2$ is introduced to make some formulas below more compact. The third order correction is zero and the terms $O(w^4)$ and smaller will be neglected.

The effective Josephson energy of the system is an integral of the effective CPR \eqref{Eq:gamma} and is given by
\begin{equation}
  U_J(\theta) = U_{J0}(\theta) + w^2 U_{J2}(\theta) + O(w^4)
  , \label{Eq:U_J(theta)}
\end{equation}
where
\begin{eqnarray}
  U_{J0}(\theta) &=& -\cos\left( \frac\kappa2 \right)\cos(\theta) - X_0 \sin\left( \frac\kappa2 \right)\sin(\theta)
  ; \label{Eq:U_J0(theta)}\\
  U_{J2}(\theta) &=& - \frac{Q}{2w^2} \sin^2\left( \frac\kappa2 \right) \cos(2\theta)
  . \label{Eq:U_J2(theta)}
\end{eqnarray}

\section{Results}
\label{Sec:Results}

\subsection{Comparison with the previous results}

For $\kappa=\pi$ we expect the result given by Eqs.~\eqref{Eq:gamma}, Eq.~\eqref{Eq:gamma0} and Eq.~\eqref{Eq:gamma2} to be similar to those previously obtained for a asymmetric 0-$\pi$ JJ\cite{Goldobin:2011:0-pi:H-tunable-CPR}. To compare both results, first, we have to convert the phases. In Ref.~\onlinecite{Goldobin:2011:0-pi:H-tunable-CPR} the phase is continuous, while in our case it is has a discontinuity. Instead of the discontinuous phase $\phi(x)$ we can introduce the continuous phase $\mu(x)$, which behaves exactly like $\phi(x)$, but without a $\kappa$-jump, \ie,
\begin{equation}
  \phi(x) = \begin{cases}
    \mu(x)          &\text{for } -w<x<x_0\\
    \mu(x) + \kappa &\text{for } x_0<x<+w
  \end{cases}
  . \label{Eq:phi(mu)}
\end{equation}
Then
\begin{equation}
  \psi\equiv\av{\phi}=\av{\mu}+(1-X_0)\frac\kappa2
  . 
\end{equation}
It is \av{\mu} that is used in Ref.~\onlinecite{Goldobin:2011:0-pi:H-tunable-CPR} (it is denoted as $\psi$ there). Rewriting our effective CPR in terms of $\av{\mu}$ and taking $\kappa=\pi$, we obtain
\begin{equation}
  \gamma = X_0 \sin\av{\mu} - Q \sin(2\av{\mu})
  . \label{Eq:gamma(mu_av)}
\end{equation}
The quantities such as $\av{j_c}$, $L_0$ and $L_\pi$ from Ref.~\onlinecite{Goldobin:2011:0-pi:H-tunable-CPR} can be expressed in terms of quantities used here as
\begin{eqnarray}
  L_0   &=& w+x_0 = w(1+X_0)
  ; \label{Eq:L0}\\
  L_\pi &=& w-x_0 = w(1-X_0)
  ; \label{Eq:Lp}\\
  \av{j_c} &=& \frac{L_0-L_\pi}{L_0+L_\pi}=X_0
  . \label{Eq:j_c_av}
\end{eqnarray}
By substituting this into expression (18) of Ref.~\onlinecite{Goldobin:2011:0-pi:H-tunable-CPR} and taking into account the definition of $\Gamma_0$, see Eq.~(17) of Ref.~\onlinecite{Goldobin:2011:0-pi:H-tunable-CPR}, we arrive at the CPR \eqref{Eq:gamma(mu_av)} derived here. Thus, for $\kappa=\pi$ the result of Ref.~\onlinecite{Goldobin:2011:0-pi:H-tunable-CPR} is reproduced \emph{exactly}. Note however that in Ref.~\onlinecite{Goldobin:2011:0-pi:H-tunable-CPR} the small parameter is the deviation of the phase from its average value, while in our case the small parameter is $w$. Although, they are related (one expects small deviations for small $w$), this relation is not straightforward.

In terms of variables used here
\begin{equation}
  \Gamma_0 = -\frac{1}{3}w^2\frac{(X_0^2-1)^2}{X_0}
  . \label{Eq:Gamma_0}
\end{equation}
If $\Gamma_0<-1$ then we have a $\pm\varphi$ JJ. This means that only for $|X_0|<w^2/3$ one obtains a $\varphi$ JJ at $\kappa=\pi$.

\subsection{Ground state phase and the critical current}
\label{Sec:Ic}

\begin{figure*}[!tb]
  \centering\includegraphics{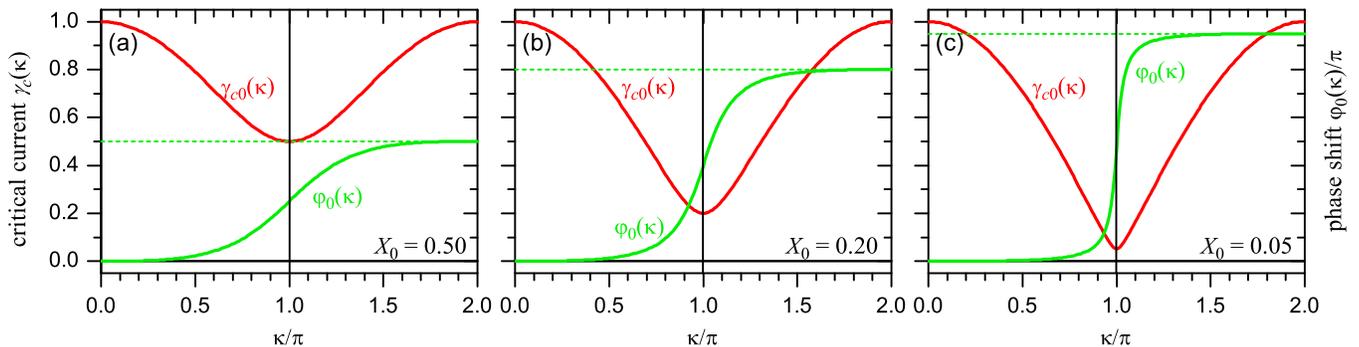}
  \caption{(Color online)
    The $\gamma_{c0}(\kappa)$ and $\varphi_0(\kappa)$ curves calculated using Eqs.~\eqref{Eq:gamma_c0(kappa)} and \eqref{Eq:varphi0(kappa)} for (a) $X_0=0.5$, (b) $X_0=0.2$ and (c) $X_0=0.05$.
  }
  \label{Fig:Ic0(kappa)}
\end{figure*}

In the 0-th approximation the averaged CPR \eqref{Eq:gamma0} can be rewritten as
\begin{equation}
  \gamma_0 = \gamma_{c0}(\kappa) \sin (\theta-\theta_0) = \gamma_{c0}(\kappa) \sin (\psi-\varphi_0)
  , \label{Eq:gamma0(psi)}
\end{equation}
where
\begin{equation}
  \gamma_{c0}(\kappa) = \sqrt{X_0^2 \sin^2\left( \frac\kappa2 \right) + \cos^2\left( \frac\kappa2 \right)}
  \label{Eq:gamma_c0(kappa)}
\end{equation}
is the maximum supercurrent. The critical current measured in experiment is $\pm\gamma_{c0}(\kappa)$. $\gamma_{c0}(\kappa)$ has maxima equal to 1 at $\kappa=2\pi n$ ($n$ is any integer) and minima equal to $X_0$ at $\kappa \bmod 2\pi =\pi$, see Fig.~\ref{Fig:Ic0(kappa)}.

In Eq.~\eqref{Eq:gamma0(psi)} the ground state phase $\varphi_0=\theta_0-X_0\kappa/2$, where
\begin{equation}
  \theta_0 = \arg\left[ \cos\left( \frac\kappa2 \right) + i\cdot X_0\sin\left( \frac\kappa2 \right)\right]
  , \label{Eq:varphi0(kappa)}
\end{equation}
and the function $\arg(z)$ returns the argument (phase angle) of a complex number $z$. Obviously, the CPR given by Eq.~\eqref{Eq:gamma0(psi)} corresponds to a $\varphi_0$ JJ\cite{Krive:2005:S-LL-S,Buzdin:2008:varphi0,Reynoso:2008:SpinPol-QPC:JosEff,Alidoust:2013:Geom-varphi0,Goldobin:2015:0-pi-SQUID=varphi-JJ,Mironov:2015:S-NW2-S}. Some examples of $\gamma_{c0}(\kappa)$ and $\varphi_0(\kappa)$ dependences are shown in Fig.~\ref{Fig:Ic0(kappa)}. For large asymmetry $X_0$ the modulation of $\gamma_{c0}(\kappa)$ is not as deep as for $X_0\to0$. The phase shift $\varphi_0(\kappa)$ changes from 0 to $\pi(\sgn(X_0)-X_0)$ as $\kappa$ changes from 0 to $2\pi$. It is positive for $X_0>0$ and negative for $X_0<0$.

When $X_0\to0$, the critical current given by the 0-th order formula \eqref{Eq:gamma_c0(kappa)} vanishes close to $\kappa=\pi$ and one has to take into account the next (second) order corrections given by Eq.~\eqref{Eq:gamma2}. This happens for $X_0 \lesssim w^2/3$, see the discussion after Eq.~\eqref{Eq:Gamma_0}.

\begin{figure*}[!tb]
  \includegraphics{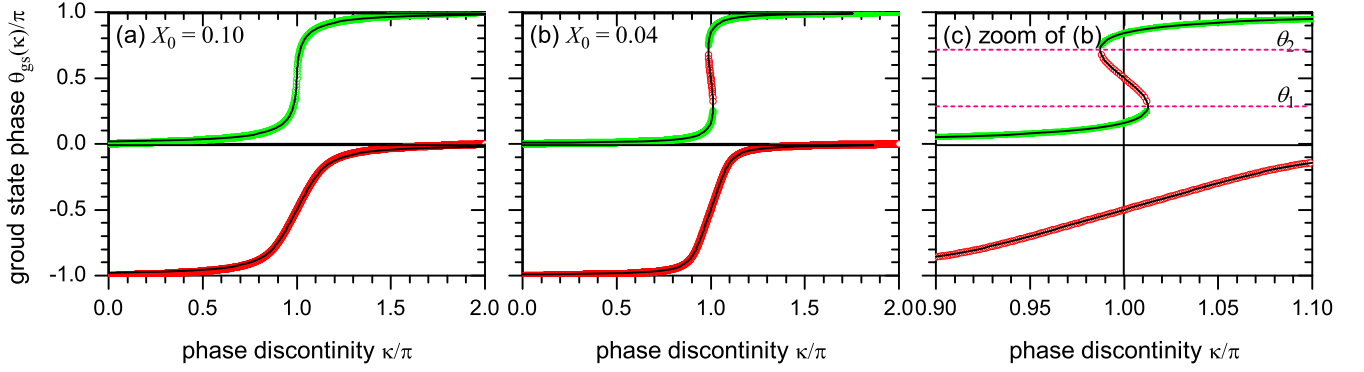}
  \caption{(Color online)
    The ground state phase $\theta_\mathrm{gs}(\kappa)$. Comparison of the approximate dependence given by Eq.~\eqref{Eq:kappa_gs(theta)} (black line) with the exact dependence calculated by numerically solving $\gamma(\theta)=0$, see Eq.~\eqref{Eq:gamma}, for each value of $\kappa$ (symbols). (a) $X_0=0.10$, (b) $X_0=0.04$, (c) $X_0=0.04$ and the region close to $\kappa=\pi$ zoomed. In all plots the regions with a positive slope (green) correspond to a stable solution (energy minimum, where $U''(\theta)=\gamma'(\theta)>0$), while the regions with a negative slope (red) correspond to an unstable one (energy maximum, where $U''(\theta)=\gamma'(\theta)<0$).
  }
  \label{Fig:GrStt}
\end{figure*}

Next, we consider the second order approximation. The ground state phase $\theta_\mathrm{gs}(\kappa)$ is a solution of $\gamma(\theta_\mathrm{gs})=0$ for this $\kappa$, \ie,
\begin{eqnarray}
  0 &=& \cos\left( \frac\kappa2 \right) \sin(\theta_\mathrm{gs})
  - X_0 \sin\left( \frac\kappa2 \right) \cos(\theta_\mathrm{gs}) +
  \nonumber\\
  &+& Q \sin^2\left( \frac\kappa2 \right) \sin(2\theta_\mathrm{gs})
  . \label{Eq:gamma=0}
\end{eqnarray}
This equation can be solved only numerically, see Fig.~\ref{Fig:GrStt}. It can be seen that multiple solutions $\theta_\mathrm{gs}(\kappa)$ appear in the vicinity of $\kappa=\pi$, see Fig.~\ref{Fig:GrStt}(b,c). To find the approximate analytical expression describing them we take $\kappa=\pi+\dk$ ($|\dk|\ll1$). Then we expand Eq.~\eqref{Eq:gamma=0}, up to the first order in $\dk$ and solve it for $\dk$. Finally we obtain an approximate value of $\kappa$ for any given ground state phase $\theta_\mathrm{gs}$,
\begin{equation}
  \kappa(\theta_\mathrm{gs}) \approx \pi + 2\left[ 2Q \cos(\theta_\mathrm{gs}) - X_0 \cot(\theta_\mathrm{gs}) \right]
  , \label{Eq:kappa_gs(theta)}
\end{equation}
\ie, the inverse of the ground state phase $\theta_\mathrm{gs}(\kappa)$. This approximation is also shown in Fig.~\ref{Fig:GrStt}. One can see that approximation given by Eq.~\eqref{Eq:kappa_gs(theta)} is very good in the whole range of $0<\kappa<2\pi$. Note, that the appearance of three solutions (two stable and one unstable) out of one near $\kappa=\pi$ is a result of the competition of the $\cos$-term with the $\cot$-term in Eq.~\eqref{Eq:kappa_gs(theta)}. From Eq.~\eqref{Eq:kappa_gs(theta)} one can figure out that the multiple solutions appear for $X_0<2Q$, \ie, $|X_0|<w^2/3$, which is in agreement with the discussion after Eq.~\eqref{Eq:Gamma_0}. We note that if $X_0$ is so small it can be neglected in the definition of $Q$, so that for $X_0 \sim w^2$ when the second order approximation becomes important, $Q=w^2/6$. If $|X_0|\gg w^2$, one can omit the $2Q\cos(\theta)$ term $\sim w^2$ in Eq.~\eqref{Eq:kappa_gs(theta)} and end up practically with expression \eqref{Eq:varphi0(kappa)} for $\theta_0$ from the 0-th order approximation. The difference is that \eqref{Eq:varphi0(kappa)} gives only the stable branch, \cf the ground state phase shown in Fig.~\ref{Fig:Ic0(kappa)} (only stable branch) and Fig.~\ref{Fig:GrStt} (both branches).

From Eq.~\eqref{Eq:kappa_gs(theta)} one can find the range of $\kappa$ where the double ground state exists, \ie, the points $\theta_1$ and $\theta_2$ in Fig.~\ref{Fig:GrStt} where $d\kappa(\theta_\mathrm{gs})/d\theta_\mathrm{gs}=0$. We obtain that
\begin{equation}
  \theta_1 = \arcsin \sqrt[3]{\frac{X_0}{2Q}}
  \text{ and }
  \theta_2 = \pi-\arcsin \sqrt[3]{\frac{X_0}{2Q}}
  , \label{Eq:theta.GrStt}
\end{equation}
which lay symmetrically with respect to $\theta=\pi/2$. It follows from Eqs.~\eqref{Eq:theta.GrStt} that the bifurcation point, where the $\theta(\kappa)$ curve switches from one stable to three solutions (two stable and one unstable), corresponds to $\theta_1=\theta_2=\pi/2$, \ie, at $X_0=2Q \approx w^2/3$, which is again in agreement with the result obtained directly from Eq.~\eqref{Eq:kappa_gs(theta)}. The range of $\kappa$ around $\kappa=\pi$ where two stable solutions exist, is found by substituting Eq.~\eqref{Eq:theta.GrStt} into Eq.~\eqref{Eq:kappa_gs(theta)}.

Looking at the Fig.~\ref{Fig:GrStt} one sees that the ground state phase $\theta_\mathrm{gs}$ has the values symmetrically placed around $\theta=(\pi/2)\sgn(X_0)$ only at $\kappa=\pi$. From Eq.~\eqref{Eq:kappa_gs(theta)} they are given by
\begin{equation}
  \theta_\mathrm{gs1}(\pi) = \arcsin\left[ \frac{X_0}{2Q} \right]
  \text{ and }
  \theta_\mathrm{gs2}(\pi) = \pi-\arcsin\left[ \frac{X_0}{2Q} \right]
  . \label{Eq:theta_gs(pi)}
\end{equation}
For $\kappa\neq\pi$ the symmetry is brocken because the corresponding double-well potential Eq.~\eqref{Eq:U_J(theta)} becomes asymmetric (one well is deeper than the other) relative to $\theta=(\pi/2)\sgn(X_0)$. The real (measurable) ground state phases are given by $\psi_\mathrm{gs}=\theta_\mathrm{gs}-X_0\kappa/2$ and at $\kappa=\pi$ are symmetric with respect to the phase $(\pi/2)[\sgn(X_0)-X_0]$. Recalling that for doubly generate state to occur one need very small $|X_0|\lesssim w^2$, the shift from $(\pi/2)\sgn(X_0)$ is small. Thus, such a JJ can be called a $\varphi_0\pm\varphi$ JJ, where, at $\kappa=\pi$ $\varphi_0=(\pi/2)[\sgn(X_0)-X_0]$ and $\varphi=\pi/2-\arcsin(X_0/2Q)$.

\begin{figure*}[!tb]
  \centering\includegraphics{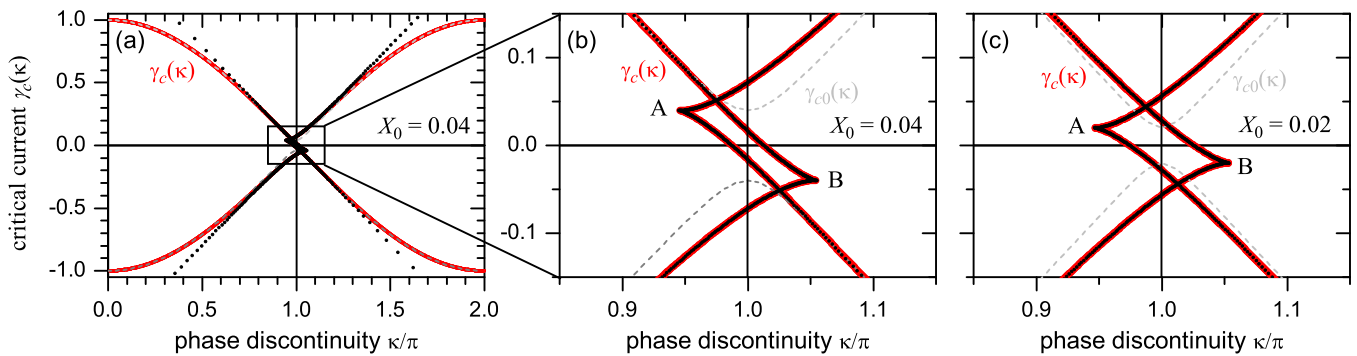}
  \caption{(Color online)
    Examples of the $\gamma_c(\kappa)$ dependence for $w=0.5$. (a) $X_0=0.04$, global behavior of $\gamma_c(\kappa)$; (b) $X_0=0.04$, zoom of the region of interest near $\kappa=\pi$; (c) $X_0=0.02$, zoom of the region near $\kappa=\pi$. Thick (red) lines/symbols show $\gamma_c$ obtained by directly solving Eq.~\eqref{Eq:gamma'(theta_c)=0} numerically to find all $\theta_c$ and then calculating $\gamma_c$ from Eq.~\eqref{Eq:gamma}. Thinner black lines/symbols correspond to the approximation given by Eq.~\eqref{Eq:gamma_c.ap(theta)}. Gray dashed lines show $\gamma_{c0}(\kappa)$, see Eq.~\eqref{Eq:gamma_c0(kappa)}, for the same parameters.
  }
  \label{Fig:Ic(kappa)}
\end{figure*}

To find the critical current(s) in the second order approximation for each $\kappa$ we search for an extremum of $\gamma(\theta,\kappa)$ with respect to $\theta$. It takes place at $\theta=\theta_c$ for which
\begin{eqnarray}
  \gamma'(\theta_c,\kappa)
  &=& X_0 \sin\left( \frac\kappa2 \right) \sin(\theta_c) + \cos\left( \frac\kappa2 \right) \cos(\theta_c) +
  \nonumber\\
  &+& 2Q \sin^2\left( \frac\kappa2 \right) [2\cos^2(\theta_c)-1] = 0
  . \label{Eq:gamma'(theta_c)=0}
\end{eqnarray}
Here and below the prime denotes $\partial/\partial\theta$. This equation can be solved for $\theta_c$ only numerically to find several (up to 4) $\theta_c$ for each value of $\kappa$. Then we substitute each of these $\theta_c$ into Eq.~\eqref{Eq:gamma} to find $\gamma_c(\kappa)=\gamma(\theta_c,\kappa)$. The result is presented in Fig.~\ref{Fig:Ic(kappa)}. The global behavior is defined mainly by $\gamma_0$, \ie, $\gamma_{c0}$. However, near $\kappa=\pi$, where $\gamma_{c0}$ vanishes, $\gamma_2$ results in a bistability and in the formation of a $\diamondsuit$-like intersection of the branches. Such $\gamma_c$ behavior is typical for a $\varphi$ JJ made of 0 and $\pi$ parts\cite{Goldobin:2011:0-pi:H-tunable-CPR,Lipman:2014:varphiEx,Goldobin:2015:0-pi-SQUID=varphi-JJ}.

Similar to the case of the ground state phase, one can find an approximate expression for $\gamma_c(\kappa)$ near $\kappa=\pi$. By substituting $\kappa=\pi+\dk$ ($|\dk|\ll1$) into Eq.~\eqref{Eq:gamma'(theta_c)=0}, Taylor-expanding it up to terms $O(\dk)$, and expressing $\dk$, we obtain the critical value of $\kappa$ corresponding to $\gamma_c$ for given $\theta$.
\begin{equation}
  \kappa_c(\theta) = \pi+\dk \approx \pi + 2\frac{X_0\sin(\theta) + 2Q\cos(2\theta)}{\cos(\theta)}
  . \label{Eq:kappa_c(theta)}
\end{equation}
To calculate the critical current, we substitute Eq.~\eqref{Eq:kappa_c(theta)} into Eq.~\eqref{Eq:gamma}, which was preliminary expanded near $\kappa=\pi$ up to $O(\dk)$. We obtain
\begin{eqnarray}
  \gamma_c^\mathrm{ap}(\theta) &\approx& -X_0\cos(\theta) + Q \sin(2\theta) -
  \nonumber\\
  &-& \frac{X_0\sin^2(\theta)+2Q\sin(\theta)\cos(2\theta)}{\cos(\theta)}
  . \label{Eq:gamma_c.ap(theta)}
\end{eqnarray}
By sweeping $\theta$ in the range $-\pi\ldots\pi$, we can now calculate $\kappa_c(\theta)$ and $\gamma_c^\mathrm{ap}(\theta)$ and make a parametric plot of $\gamma_c^\mathrm{ap}(\theta)$ \vs $\kappa_c(\theta)$, see Fig.~\ref{Fig:Ic(kappa)}. The agreement with direct numerical calculations near $\kappa=\pi$ is excellent, see Fig.~\ref{Fig:Ic(kappa)}(b,c). The deviations become noticeable as $\kappa \bmod 2\pi$ approaches 0 or $2\pi$, see Fig.~\ref{Fig:Ic(kappa)}(a).

This approximate analytical expression for $\gamma_c$ allows us to calculate some key features in the $\gamma_c(\kappa)$ plot. For example, one can find out the value of $\theta_\beak$ (and $\kappa_\beak$), for which the branches meet each other, see points A and B in Fig.~\ref{Fig:Ic(kappa)}. The analysis of the $\kappa_c(\theta)$ dependence \eqref{Eq:kappa_c(theta)} shows that this happens when $d\kappa_c/d\theta=0$. Differentiating Eq.~\eqref{Eq:kappa_c(theta)} we obtain the following equation for $\theta_\beak$.
\begin{equation}
  4Q\sin^3(\theta_\beak) - 6Q\sin(\theta_\beak) + X_0 = 0
  . \label{Eq:sin(theta_beak)}
\end{equation}
This cubic equation with respect to $\sin(\theta_\beak)$ has only one suitable root, which (after some lengthy algebra) can be expressed as
\begin{equation}
  \sin(\theta_\beak) = \frac{-1}{\sqrt{2}}\left[ \cos\left( \frac\chi3 \right) - \sqrt{3} \sin\left( \frac\chi3 \right) \right]
  , \label{Eq:theta_beak}
\end{equation}
where $\chi$ may be explicitely written as
\begin{equation}
  \chi = \arg\left[ -X_0 + i\cdot\sqrt{8Q^2-X_0^2} \right]
  . \label{Eq:chi.def}
\end{equation}
It turns out that for typical parameters corresponding to a bistable case (small $|X_0|<w^2/3 \approx 2Q$), the value of $\chi$ changes from $\pi/4$ to $3\pi/4$. Then by expanding Eq.~\eqref{Eq:chi.def} as
\begin{equation}
  \chi \approx \frac\pi2+\frac{\sqrt{2}X_0}{4Q} \approx \frac\pi2+\frac{3}{\sqrt{2}}\frac{X_0}{w^2}
  , \label{Eq:chi.approx}
\end{equation}
we get $\sin(\theta_\beak)\approx X_0/(6Q)=X_0/w^2<1/3$. In fact, this limit of small $\sin(\theta_\beak)$ corresponds to neglecting the $\sin^3(\theta_\beak)$ term in Eq.~\eqref{Eq:sin(theta_beak)}, so that one obtains $\sin(\theta_\beak)\approx X_0/(6Q)$ right away from Eq.~\eqref{Eq:sin(theta_beak)}. After finding $\sin(\theta_\beak)$, the value of $\kappa_\beak$ can be found as $\kappa_c(\theta_\beak)$ from Eq.~\eqref{Eq:kappa_c(theta)}. Using the approximation \eqref{Eq:chi.approx}, \ie, in the worst case neglecting $4\sin^3(\theta_\beak)=4/27$ in comparison with $6\sin(\theta_\beak)=2$ (accuracy $\sim8\%$) in Eq.~\eqref{Eq:sin(theta_beak)}, we can write
\begin{equation}
  \kappa_\beak \approx \pi\pm \left[ \frac23 w^2 + \frac{X_0^2}{w^2} \right]
  . \label{Eq:kappa_beak}
\end{equation}

\subsection{Energy barrier}

\newcommand{\dth}{\ensuremath{\delta\theta}\xspace}
\newcommand{\dg}{\ensuremath{\delta\gamma}\xspace}

We consider the thermal escape or the quantum tunneling of the phase $\theta$ out of the potential well, when the bias current $\gamma\to\gamma_c(\kappa)$. Since our model reduces the system to an effective point-like JJ, for calculation of the escape rate $\Gamma$ one can use standard thermal or quasi-classical quantum formulas. In these formulas, the key parameters are the barrier height $\Delta U(\gamma)$ and the eigenfrequency $\omega_0(\gamma)$. The aim of this section is to obtain the expressions for them.

In general, we proceed as follows. Given the Josephson energy profile $U_J(\theta)$, the total potential energy of the biased JJ can be written as a tilted potential
\begin{equation}
  U(\theta) = U_J(\theta)-\gamma\theta
  . \label{Eq:U}
\end{equation}
The static solution(s) correspond(s) to
\begin{equation}
  U_J'(\theta)-\gamma = 0
  , \label{Eq:U'}
\end{equation}
In essence this is a CPR. The critical current is reached for $\theta=\theta_c$, when $d\gamma/d\theta=0$, \ie,
\begin{equation}
  U_J''(\theta_c) = 0
  . \label{Eq:U''}
\end{equation}
From here one can, in principle, find (one or more) values of $\theta_c$. Imagine that we have found all values of $\theta_c$. Then, the value of the critical current $\gamma_c$ is found from Eq.~\eqref{Eq:U'}, \ie,
\begin{equation}
  \gamma_c = U_J'(\theta_c)
  . \label{Eq:U:gamma_c}
\end{equation}
Now we assume that the value of $\gamma$ is slightly undercritical, \ie, $\gamma=\gamma_c(1-\dg)$, where $0\leq\dg\ll 1$ for any sign of $\gamma_c$. We are interested to expand the potential $U(\theta)$ in the vicinity of the bending point $\theta_c$. We, therefore, write $\theta=\theta_c+\delta\theta$ ($|\delta\theta|\ll1$) and substitute this into Eq.~\eqref{Eq:U'} and Taylor-expand up to $O(\dth^2)$. We get
\begin{equation}
  U_J'(\theta_c) + U_J''(\theta_c)\dth + \frac12 U_J'''(\theta_c)\dth^2 = \gamma_c (1-\dg)
  . \label{Eq:CPR.undercritical}
\end{equation}
Here the first terms in the \lhs and the \rhs cancel because of Eq.~\eqref{Eq:U:gamma_c}, the second term in the \lhs vanishes because of Eq.~\eqref{Eq:U''}. As a result we obtain new static solutions for an undercritical $\gamma$ shifted from $\theta_c$ by
\begin{equation}
  \dth = \sqrt{\frac{-2\gamma_c\dg}{U_J'''(\theta_c)}}
  . \label{Eq:dth}
\end{equation}
to the positive or negative direction. One of them is stable and corresponds to the minimum of $U(\theta)$, another, unstable one, corresponds to the maximum of $U(\theta)$. The energy barrier
\begin{equation}
  \Delta U = |U(\theta_c+\dth)-U(\theta_c-\dth) |
  . \label{Eq:DeltaU.def}
\end{equation}
After expanding up to $O(\dth^3)$, we see that the terms $O(\dth^0)$ and $O(\dth^2)$ cancel, and we obtain
\begin{equation}
  \Delta U = |2U_J'(\gamma_c)\dth -2\gamma\dth + \frac26 U_J'''(\theta_c)\dth^3|
  . \label{Eq:DeltaU.i1}
\end{equation}
Using Eq.~\eqref{Eq:U:gamma_c}, the definition of $\gamma=\gamma_c(1-\dg)$ and the expression \eqref{Eq:dth}, we finally obtain
\begin{equation}
  \Delta U = \frac{4\sqrt{2}}{3}\frac{|\gamma_c\dg|^{3/2}}{\sqrt{|U_J'''(\theta_c)|}}
  = \frac{4\sqrt{2}}{3}\sqrt{\left| \frac{\gamma_c^3}{U_J'''(\theta_c)} \right|} \dg^{3/2}
  . \label{Eq:DeltaU.fin.gen}
\end{equation}
Now let us apply this general result to our system.

\begin{figure*}[!ptb]
  \centering\includegraphics{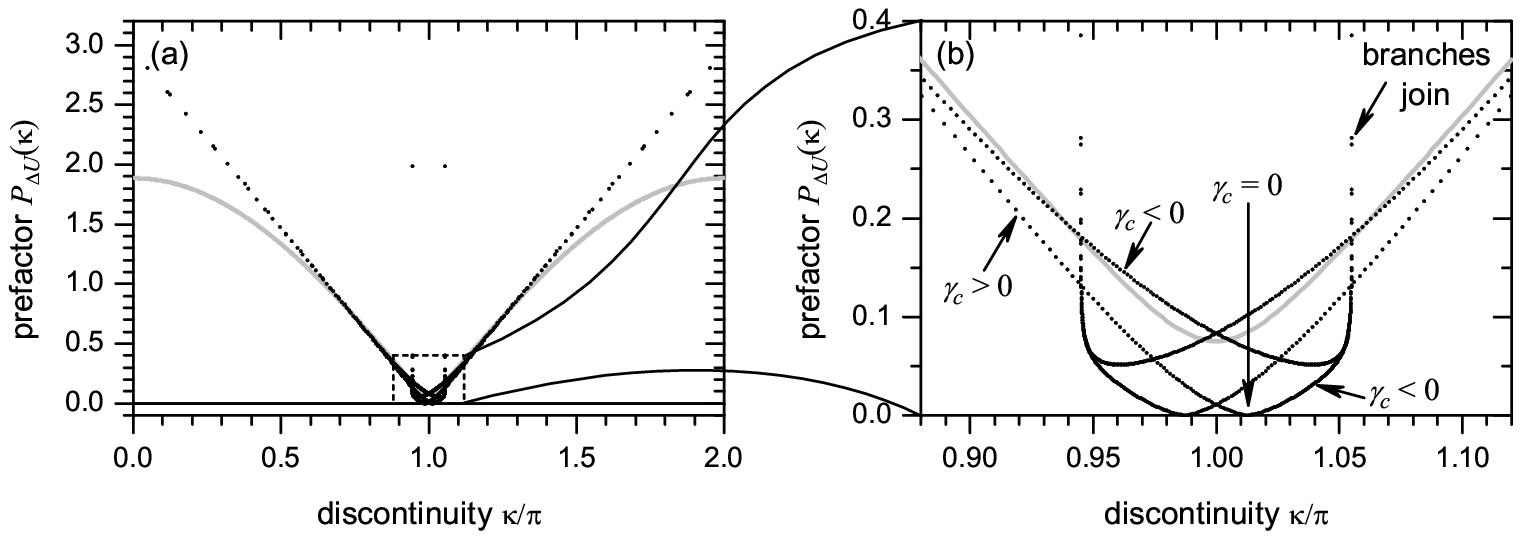}
  \caption{%
    The energy barrier prefactor $P_{\Delta U}$ given by Eq.~\eqref{Eq:2:PreDeltaU} \vs $\kappa_c(\theta)$ (black) and the one calculated using the 0-th order approximation Eq.~\eqref{Eq:0:DeltaU} (gray). (a) shows global behavior in the interval $0\leq\kappa\leq2\pi$, while (b) shows the zoom of the area close to $\kappa=\pi$, where multiple solutions appear. Parameters are: $w=0.5$, $X_0=0.04$.
  }
  \label{Fig:DeltaU}
\end{figure*}

If $X_0\gg w^2/3$ or if $\kappa$ is far away from $\kappa=\pi$, then we can use only the 0-th order term in Eq.~\eqref{Eq:gamma} and in Eq.~\eqref{Eq:U_J(theta)}. In this limit the CPR is sinusoidal, see Eq.~\eqref{Eq:gamma0(psi)}. Although it is shifted by $\theta_0$ ($\varphi_0$), it is irrelevant for calculation of the escape barrier and eigenfrequency. Thus, the system behaves as a conventional JJ with sinusoidal CPR, critical current $\gamma_{c0}(\kappa)$ and Josephson energy $U_J=\gamma_{c0}(\kappa)[1-\cos(\theta-\theta_0)]$. Thus, $\theta_c=\pi/2$,  $U'''(\theta_c)=-\gamma_{c0}$ and we obtain the usual approximation for energy barrier and eigenfrequency in the limit $\gamma\to\pm\gamma_{c0}$:
\begin{equation}
  \Delta U(\dg,\kappa) = \frac{4\sqrt{2}}{3} \gamma_{c0}(\kappa) \, \dg^{3/2}
  . \label{Eq:0:DeltaU}
\end{equation}

In the case when the second order correction is important, \ie, $X_0 \lesssim w^2/3$ and $\kappa\approx\pi$, we use the same approach, but again, like in the section about critical current and ground state phase, we approximate for $\kappa=\pi+\dk$ ($|\dk|\ll1$). In this case the energy is given by
\begin{equation}
  U(\theta, \dk) = \frac{\dk}{2}\cos(\theta) - X_0 \sin(\theta) - \frac{Q}{2}\cos(2\theta)
  . \label{Eq:U.approx}
\end{equation}
From here
\begin{equation}
  U'''(\theta, \kappa) = \frac{\kappa-\pi}{2}\sin(\theta) + X_0 \cos(\theta) - 4 Q \sin(2\theta)
  . \label{Eq:2:U'''.approx}
\end{equation}

We have to take $\theta=\theta_c$ in Eq.~\eqref{Eq:2:U'''.approx} and substitute this into Eq.~\eqref{Eq:DeltaU.fin.gen}. The dependence of $\Delta U$ on $\dg$ is obvious from Eq.~\eqref{Eq:DeltaU.fin.gen}, so our aim is to see how prefactor in Eq.~\eqref{Eq:DeltaU.fin.gen} depends on $\kappa$ (or $\dk$). Since, $\theta_c$ for each $\kappa$ can be found only numerically, we, as in the previous sections, sweep $\theta$ from $-\pi$ to $+\pi$ and find the corresponding $\kappa_c(\theta)$ from Eq.~\eqref{Eq:kappa_c(theta)} and then calculate $U'''(\theta,\kappa_c(\theta))$ from Eq.~\eqref{Eq:2:U'''.approx}. Then we make a parametric plot of the energy barrier prefactor
\begin{equation}
  P_{\Delta U}(\theta) \equiv \frac{\Delta U(\theta,\kappa_c(\theta))}{\dg^{3/2}}
  = \frac{4\sqrt2}{3} \sqrt{\left| \frac{\gamma_c(\theta,\kappa_c(\theta))^3}{U'''(\theta,\kappa_c(\theta))} \right|}
  , \label{Eq:2:PreDeltaU}
\end{equation}
as a function of $\kappa_c(\theta)$, see Fig.~\ref{Fig:DeltaU}. The global behavior is given by the 0-th order approximation, see prefactor in Eq.~\eqref{Eq:0:DeltaU}. The second order approximation, where we expanded all expressions near $\kappa=\pi$, works well near $\kappa=\pi$, but deviates substantially from the real solution given by the 0-th order approximation when $(\kappa \bmod 2\pi)\to 0$ or $\to2\pi$. In Fig.~\ref{Fig:DeltaU}(b), as $\kappa$ increases, one sees two branches, given by Eq.~\eqref{Eq:2:PreDeltaU}. One of them corresponds to the negative critical current branch, \cf Fig.~\ref{Fig:Ic(kappa)}(b), another to the positive one. At $\kappa$ slightly larger than $\pi$ the positive $\gamma_c(\kappa)$ branch crosses zero, see Fig.~\ref{Fig:Ic(kappa)}(b), so that we see that the prefactor also vanishes at this point, see Fig.~\ref{Fig:DeltaU}(b). Then, at somewhat larger $\kappa$ both mentioned branches join, see  Fig.~\ref{Fig:Ic(kappa)}(b). At this point the prefactor diverges, see Fig.~\ref{Fig:DeltaU}(b). The other two branches in Fig.~\ref{Fig:DeltaU}(b) show similar behavior.

\subsection{Eigenfrequency}

\begin{figure*}[!ptb]
  \centering\includegraphics{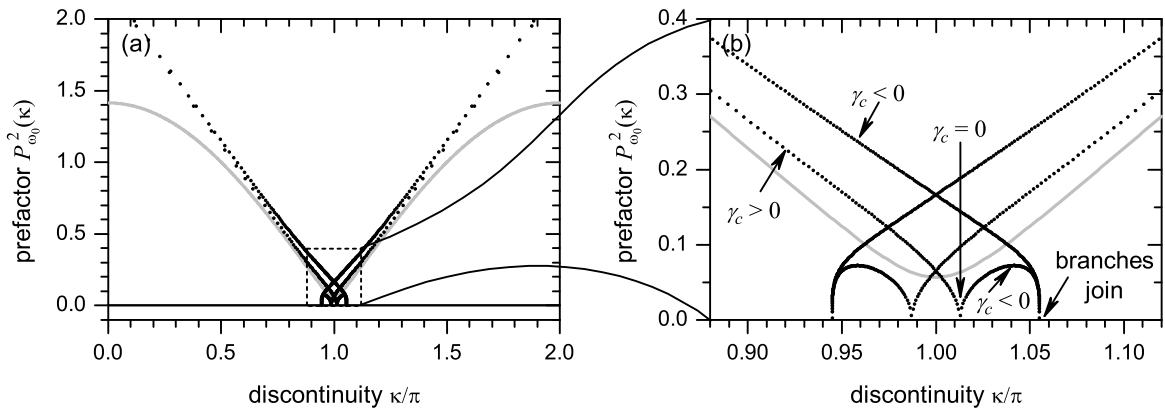}
  \caption{
    The eigenfrequency prefactor $P_{\omega_0}^2$ given by Eq.~\eqref{Eq:2:PreEigenFreqSqr} \vs $\kappa_c(\theta)$ given by Eq.~\eqref{Eq:kappa_c(theta)} (black) and the one calculated using the 0-th order approximation Eq.~\eqref{Eq:0:EigenFreq} (gray). (a) shows global behavior in the interval $0\leq\kappa\leq2\pi$, while (b) shows the zoom of the area close to $\kappa=\pi$, where multiple solutions appear. Parameters are: $w=0.5$, $X_0=0.04$.
  }
  \label{Fig:PreEigenFreqSqr}
\end{figure*}

In general, the eigenfrequency of phase oscillations around one of the static solution $\theta_c\pm\dth$, see Eq.~\eqref{Eq:dth}, is given by
\begin{equation}
  \omega_0^2 = U''(\theta_c\pm\dth) = U''(\theta_c) \pm U'''(\theta_c) \dth
  . \label{Eq:omega_0.def}
\end{equation}
The first term vanishes because of Eq.~\eqref{Eq:U''}. So, using Eq.~\eqref{Eq:dth} we get
\begin{equation}
  \omega_0^2 = \sqrt{|2\gamma_c U'''(\theta_c)|} \,\dg^{1/2}
  . \label{Eq:EigenFreq.gen.final}
\end{equation}

In the 0-th approximation $\theta_c=\pi/2$, $U'''(\theta_c)=-\gamma_{c0}$ and we arrive to the well-known result
\begin{equation}
  \omega_0^2 = \gamma_{c0} \sqrt{2} \, \dg^{1/2}
  . \label{Eq:0:EigenFreq}
\end{equation}

In the second order approximation we again sweep $\theta$ and make an implicit plot of the $\omega_0^2$ prefactor
\begin{equation}
  P_{\omega_0}^2(\theta) \equiv \omega_0^2/\dg^{1/2}
  = \sqrt{|2\gamma_c(\theta,\kappa_c(\theta))U'''(\theta,\kappa_c(\theta))|}
  , \label{Eq:2:PreEigenFreqSqr}
\end{equation}
as a function of $\kappa_c(\theta)$ given by Eq.~\eqref{Eq:kappa_c(theta)}. The behavior of the eigenfrequency prefactor is shown in Fig.~\ref{Fig:PreEigenFreqSqr}. Similar to the energy barrier prefactor, the eigenfrequency prefactor given by Eq.~\eqref{Eq:2:PreEigenFreqSqr} describes the multiple solutions near $\kappa=\pi$ well. However,  the 0-th approximation is better outside this vicinity.

\subsection{Escape histogram width in the MQT regime}

\begin{figure*}[!ptb]
  \centering\includegraphics{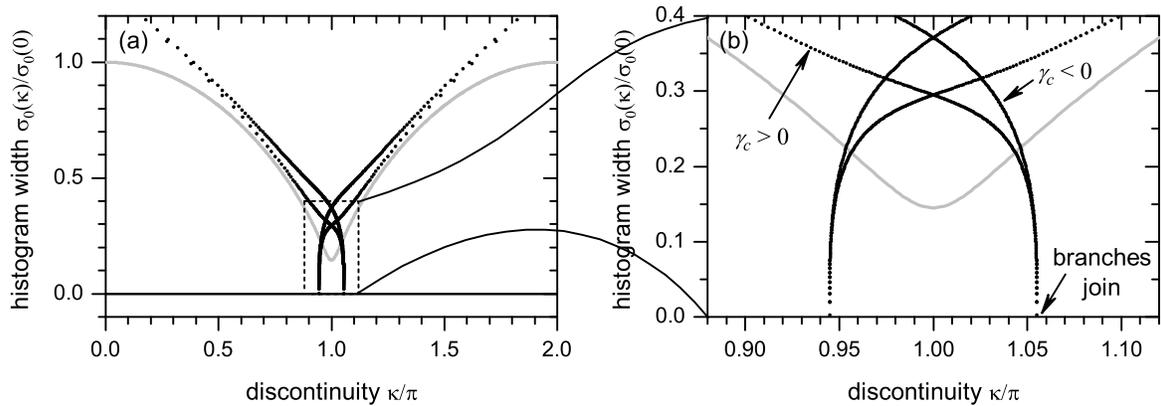}
  \caption{%
    The width $\sigma(\kappa)$ of the escape current histogram in MQT regime calculated using the 0-th order approximation Eq.~\eqref{Eq:sigma0} (gray line) and the second order approximation linearized near $\kappa=\pi$, see Eq.~\eqref{Eq:sigma}. (a) shows the global behavior in the interval $0\leq\kappa\leq2\pi$, while (b) shows the zoom of the area close to $\kappa=\pi$, where multiple solutions appear. Parameters are: $w=0.5$, $X_0=0.04$.
  }
  \label{Fig:sigma(kappa)}
\end{figure*}

The dependences $\Delta U(\dg,\kappa)$ and $\omega_0(\dg,\kappa)$ allow to directly calculate not only the escape rate, but also the width $\sigma$ of the escape histogram as a function of $\kappa$. This $\sigma(\kappa)$ dependence can be directly compared with the experimentally measured one. For the sake of simplicity we limit ourselves to the case of MQT, so that the temperature is excluded. The approximate, but rather precise, formula for the histogram width (dispersion) $\sigma$ was derived by Garg\cite{Garg:1995:EscapeDistr} in the general case of a particle in a tilted potential. For MQT regime the Garg\cite{Garg:1995:EscapeDistr} expression reduces to
\begin{equation}
  \sigma_{\dg} \propto \left[ \frac{P_{\omega_0}}{P_{\Delta U}} \right]^\frac45
  , \label{Eq:sigma_norm.gen}
\end{equation}
where we have omitted $\ln$-terms that are much weaker than power-terms. We use a $\propto$ sign as we are interested not in the width itself but in its scaling as a function of $\kappa$. This $\sigma$ is a dispersion of $\dg$ defined above, \ie, it assumes that the critical current is equal to 1. If the critical current is equal to $\gamma_c$, the sigma (measured in the same units as $\gamma_c$) is
\begin{equation}
  \sigma(\kappa) = \left[ \frac{P_{\omega_0(\kappa)}}{P_{\Delta U(\kappa)}} \right]^\frac45 |\gamma_c(\kappa)|
  . \label{Eq:sigma}
\end{equation}

When $X_0\gg w^2/3$, or $\kappa \bmod 2\pi$ is not very close to $\pi$, we can use the 0-th order approximation. In this case, by substituting the prefactors from Eq.~\eqref{Eq:0:DeltaU} and Eq.~\eqref{Eq:0:EigenFreq} into Eq.~\eqref{Eq:sigma}, we get
\begin{equation}
  \sigma_0(\kappa) \propto \gamma_{c0}^{3/5}(\kappa)
  . \label{Eq:sigma0}
\end{equation}
This dependence is shown by the gray line in Fig.~\ref{Fig:sigma(kappa)}.

For small $X_0\lesssim w^2/3$ and $\kappa \bmod 2\pi$ in the vicinity of $\pi$, we have to use the second order formulas. Again by substituting prefactors from Eq.~\eqref{Eq:2:PreDeltaU} and Eq.~\eqref{Eq:2:PreEigenFreqSqr} into Eq.~\eqref{Eq:sigma} we obtain $\sigma(\theta)$, which we plot \vs $\kappa_c(\theta)$ as a parametric plot, see Fig.~\ref{Fig:sigma(kappa)}. One can see that $\sigma$ vanishes at the bifurcation point where the two branches join. Note also that at the points $\kappa=\kappa_z$ where $\gamma_c(\kappa)$ vanishes (crosses zero), \ie $\gamma_c\propto\kappa_z+\dk$ is linear, both $P_{\Delta U}\propto\dk^{3/2}$ and $P_{\omega_0}\propto\dk^{1/4}$ vanish, however $\sigma\propto \const$ does not have zero or any other peculiarity at these points, as can be seen from Eq.~\eqref{Eq:sigma}.

\subsection{Experimental relevance}

On one hand the range of $|X_0|<w^2/3$ ($|x_0|<w^3/3$), required to create a $\varphi$ JJ near $\kappa=\pi$, is very tiny. This was already pointed out in the previous works\cite{Buzdin:2003:phi-LJJ,Goldobin:2011:0-pi:H-tunable-CPR,Sickinger:2012:varphiExp,Lipman:2014:varphiEx,Goldobin:2015:0-pi-SQUID=varphi-JJ}. This makes it very difficult to controllably fabricate the desired $x_0$ --- a small technological shift can drastically change the junction.

However even a nominally symmetric junction ($X_0=0$), due to a tiny technological misalignment can get $X_0\neq 0$. As a result an experimental $|\gamma_c(\kappa)|$ curve exhibits asymmetric minima for positive and negative bias current, as in Fig.~\ref{Fig:Ic(kappa)}. Also, in experiment it should be easy to conclude whether the asymmetry $|X_0|$ is smaller than $w^2/3$ (and we deal with a $\pm\varphi$ JJ) or larger (and we deal with single state $\varphi_0$ JJ). In the former case the $\gamma_c(\kappa)$ dependence should have a cusp-like minimum with branch crossing, while in the latter case the minimum will be smooth.

\section{Conclusions}
\label{Sec:Conclusions}

We have derived an effective model, which describes a short JJ with a phase discontinuity $\kappa$ at an arbitrary point $x_0$ along its length. This model reduces the system considered here to a point-like JJ with an unconventional CPR. One can relatively easy obtain all desired characteristics of such a point-like JJ. For example, we analyzed the ground state and found that close to $\kappa=\pi$ one obtains a $\varphi_0\pm\varphi$ JJ, while far from this point it is $\varphi_0$ JJ. We also calculated the dependence of the critical current of such a JJ as a function of $\kappa$ and found multiple branches close to $\kappa=\pi$, corresponding to $\varphi_0\pm\varphi$ states. Further, we have calculated the behavior of the energy barrier and eigenfrequency close to the critical current, which allow to make estimations of the width $\sigma$ of the switching current histogram in the regime of macroscopic quantum tunneling.

\appendix
\section{Derivation of the averaged CPR}
\label{Sec:Derivation}

We use the following phase ansatz
\begin{widetext}
\begin{equation}
  \phi(x) = \phi_0 + \begin{cases}
     -\frac{\kappa}{2} + A_L (x-x_0) + \frac12 B_L(x-x_0)^2 + \frac13C_L(x-x_0)^3 + \frac14 D_L (x-x_0)^4, & -w<x<x_0\\
     +\frac{\kappa}{2} + A_R (x-x_0) + \frac12 B_R(x-x_0)^2 + \frac13C_R(x-x_0)^3 + \frac14 D_R (x-x_0)^4, & x_0<x<+w
   \end{cases}
   , \label{Eq:ansatz}
\end{equation}
\end{widetext}
which corresponds to the phase discontinuity at $x=x_0$ and a Taylor expansion of the phase $\phi(x)$ in the left and right region (subscripts $L$ and $R$) relative to the discontinuity $x_0$.

In statics, the phase $\phi(x)$ should satisfy the Ferrel-Prange equation
\begin{equation}
  \gamma= \sin(\phi) - \phi''
  , \label{Eq:FP}
\end{equation}
subject to boundary conditions at the edges $x=\pm w$, corresponding to zero applied magnetic field, and the field continuity at $x=x_0$:
\begin{subequations}
  \begin{eqnarray}
    \phi'(-w) &=& 0
    ; \label{Eq:BC@-w}\\
    \phi'(+w) &=& 0
    ; \label{Eq:BC@+w}\\
    \phi'(x_0-0) &=& \phi'(x_0+0)
    . \label{Eq:BC@x0}
  \end{eqnarray}
  \label{Eq:BCs}
\end{subequations}
The prime denotes $\partial/\partial x$. Below, we use the junction half-length $w$ as a small parameter and develop a perturbation theory with respect to $w$.

\subsection{0-th approximation in $w$}

We substitute the ansatz Eq.~\eqref{Eq:ansatz} into the Ferrel-Prange Eq.~\eqref{Eq:FP}. After calculating $\phi''$, we would like to expand $\sin(\phi)$ with respect to the small parameter $w$. The key point is to make this expansion correctly. For this, we transform the argument of the sine function to explicitly pull out $w$ from all terms. Namely, we define that $A_{L,R}$ and $B_{L,R}$ from ansatz Eq.~\eqref{Eq:ansatz} depend on $w$ as
\begin{eqnarray}
  x-x_0 &=& \dX w
  ; \label{Eq:dX}\\
  A_{L,R} &=& a_{L1,R1} w
  ; \label{Eq:A_LR}\\
  B_{L,R} &=& b_{L0,R0}
  . \label{Eq:B_LR}
\end{eqnarray}
Here and below the subscripts separated by a comma mean that it is actually two equations: one is obtained by taking the first subscript in the whole equation, the second equation is obtained by taking the second subscript in the whole equation.
The higher order $C$ and $D$-terms from Eq.~\eqref{Eq:ansatz} are not relevant in $0$-th approximation. Initially, the scaling of $A$ and $B$ with $w$ is actually not obvious, but later we will see that the scaling given by Eqs.~\eqref{Eq:A_LR} and \eqref{Eq:B_LR} is consistent. After the above substitution we expand $\sin(\ldots)$ in Eq.~\eqref{Eq:FP} relative to $w$, keeping only constant terms (neglecting $O(w)$ and smaller). We arrive at the following expression(s).
\begin{eqnarray}
  b_{L0,R0} &=& \sin\left( \phi_0 \mp \frac\kappa2 \right) - \gamma_0
  , \label{Eq:b_LR0}
\end{eqnarray}
where $\gamma=\gamma_0$ in our $0$-th approximation. From the Eqs.~\eqref{Eq:b_LR0} it is obvious that $B$ scales $\sim w^0$ as written in Eq.~\eqref{Eq:B_LR}. From the boundary conditions Eq.~\eqref{Eq:BCs} we have
\begin{subequations}
  \begin{eqnarray}
    a_{L1} &=& (X_0+1) b_{L0}
    ; \label{Eq:0:BC@-w}\\
    a_{R1} &=& (X_0-1) b_{L0}
    ; \label{Eq:0:BC@+w}\\
    a_{L1} &=& a_{R1}
    , \label{Eq:0:BC@x0}
  \end{eqnarray}
\end{subequations}
where $X_0=x_0/w$. It is Eqs.~\eqref{Eq:0:BC@-w} and \eqref{Eq:0:BC@+w} where it becomes obvious that $A_{L,R}\sim w$, as it was correctly written in Eq.~\eqref{Eq:A_LR}, otherwise the \lhs and the \rhs would have different orders in $w$. By substituting $b_{L0,R0}$ from Eqs.~\eqref{Eq:b_LR0} into Eqs.~\eqref{Eq:0:BC@-w} and \eqref{Eq:0:BC@+w} and then $a_{L1}$ and $a_{R1}$ from Eqs.~\eqref{Eq:0:BC@-w} and \eqref{Eq:0:BC@+w} into Eq.~\eqref{Eq:0:BC@x0}, we finally get the current-phase relation
\begin{equation}
  \gamma_0(\phi_0) = \cos\left( \frac\kappa2 \right)\sin(\phi_0) -X_0 \sin\left( \frac\kappa2 \right) \cos(\phi_0)
  . \label{Eq:0:CPR}
\end{equation}

\subsection{2-nd order approximation}

In the next order ($\sim w^2$) approximation we use all the terms in ansatz \eqref{Eq:ansatz} to substitute into the Eq.~\eqref{Eq:FP}. After calculating $\phi''$, we explicitly extract $w$ from all terms using the following substitutions
\begin{subequations}
  \begin{eqnarray}
    A_{L,R} &=& a_{L1,R1} w + a_{L3,R3}w^3
    ; \label{Eq:2:A_LR}\\
    B_{L,R} &=& b_{L0,R0} + b_{L2,R2} w^2
    ; \label{Eq:2:B_LR}\\
    C_{L,R} &=& c_{L1,R1}w
    ; \label{Eq:2:C_LR}\\
    \gamma &=& \gamma_0 + w^2 \gamma_2
    . \label{Eq:2:gamma}
  \end{eqnarray}
  \label{Eq:ABCD(w)}
\end{subequations}
Here $a_{L1,R1}$, $b_{L0,R0}$ and $\gamma_0$ are from the 0-th order approximation and $a_{L3,R3}$ and $b_{L2,R2}$ are the next order corrections. Other powers, \eg, $a_{L2,R2}=b_{L1,R1}=0$. After the above substitution we expand $\sin(\ldots)$ in Eq.~\eqref{Eq:FP} relative to $w$, keeping only terms $\sim O(w^2)$ and larger (neglecting $O(w^3)$ and smaller). We arrive at the second order polynomial in $\dX$ equal to zero. Obviously, it can be equal to zero for any $\dX$ only if each coefficient in front of $\dX^2$, $\dX$ and constant are all equal to zero. We, thus, obtain
\begin{subequations}
  \begin{eqnarray}
    \dX^0 &:& b_{L2,R2} = -\gamma_2
    ; \label{Eq:2:gamma2.def}\\
    \dX^1 &:& c_{L1,R1} = \frac12 a_{L1,R1} \cos\left( \phi_0 \mp \frac\kappa2 \right)
    ; \label{Eq:2:c_L}\\
    \dX^2 &:& D_{L,R} = \frac16 b_{L0,R0} \cos\left( \phi_0 \mp \frac\kappa2 \right)
    . \label{Eq:2:D_L}
  \end{eqnarray}
  \label{Eq:2:bcD}
\end{subequations}

As a next step we substitute the ansatz \eqref{Eq:ansatz} into the boundary conditions \eqref{Eq:BCs}. Then we substitute the definitions \eqref{Eq:ABCD(w)}, cancel the terms from 0-th order approximation (if any), substitute $c_{L1,R1}$, $D_{L,R}$ and $b_{L2,R2}$ from Eqs.~\eqref{Eq:2:bcD} and obtain
\begin{widetext}
\begin{subequations}
  \begin{eqnarray}
    a_{L3} &=& -\frac{b_{L0}}3 \cos\left( \phi_0 - \frac\kappa2 \right)  (X_0+1)^3 -\gamma_2 (X_0+1)
    ; \label{Eq:2:BC@-w}\\
    a_{R3} &=& -\frac{b_{R0}}3 \cos\left( \phi_0 + \frac\kappa2 \right) (X_0-1)^3 -\gamma_2 (X_0-1)
    ; \label{Eq:2:BC@+w}\\
    a_{L3} &=& a_{R3}
    . \label{Eq:2:BC@x0}
  \end{eqnarray}
  \label{Eq:2:BCs}
\end{subequations}
\end{widetext}
By substituting Eqs.~\eqref{Eq:2:BC@-w} and \eqref{Eq:2:BC@+w} into Eq.~\eqref{Eq:2:BC@x0} we finally obtain
\begin{equation}
  \gamma_2 = \frac16 \cos\left( \phi_0 + \frac\kappa2 \right) b_{R0} - \frac16 \cos\left( \phi_0 - \frac\kappa2 \right) b_{L0}
  . \label{Eq:2:gamma2(b_0)}
\end{equation}
Finally, we substitute $b_{L0,R0}$ from Eq.~\eqref{Eq:b_LR0} (0-th approximation) and obtain the final expression for the second order correction to the current
\begin{eqnarray}
  \gamma_2(\phi_0)
  &=& \frac16 (1-X_0^4) \sin^2\left( \frac\kappa2 \right)  \sin(2\phi_0) +
  \nonumber\\
  &+& \frac16 X_0 (1-X_0^2) \sin(\kappa) [1+\cos(2\phi_0)]
  , \label{Eq:2:CPR}
\end{eqnarray}

\subsection{From $\phi_0$ to the average phase $\psi$}

Up to now both the 0-th order CPR Eq.~\eqref{Eq:0:CPR} and the second order CPR Eq.~\eqref{Eq:0:CPR} are given as functions of the $\phi_0$, while our aim is to express those as a function of the average phase $\psi$. To find $\psi$, we substitute ansatz Eq.~\eqref{Eq:ansatz} into Eq.~\eqref{Eq:psi.def}, integrate on each interval, substitute the definitions of $A$, $B$, $C$, $D$ from Eqs.~\eqref{Eq:ABCD(w)}, then substitute expressions for $a$, $b$, $c$, $D$ and $\gamma_0$ from Eqs.~\eqref{Eq:b_LR0}, \eqref{Eq:2:bcD}, \eqref{Eq:2:BCs} and \eqref{Eq:0:CPR}. Then we keep only the terms $\sim O(w^2)$ and larger, after some simplifications arrive at
\begin{equation}
  \psi = \phi_0 - \frac\kappa2 X_0 +w^2\frac23 X_0 (1-X_0^2) \sin\left( \frac\kappa2 \right) \cos(\phi_0)
  . \label{Eq:psi(phi0)}
\end{equation}
Our aim is to invert this expression, \ie, to express $\phi_0(\psi)$ to substitute to CPR and obtain $\gamma(\psi)$. We again act following the perturbation theory with respect to the small parameter $w$. In the 0-th approximation
\begin{equation}
  \phi_0^{(0)} = \psi + \frac\kappa2 X_0 \equiv \theta
  . 
\end{equation}
Here we introduced the angle $\theta$, which makes expressions more compact.

In the next (second) approximation $\phi_0 = \phi_0^{(0)} + w^2 \phi_0^{(2)}$. By substituting this into Eq.~\eqref{Eq:psi(phi0)} and expanding up to $O(w^2)$, we obtain $\phi_0^{(2)}$ and therefore
\begin{equation}
  \phi_0(\psi) = \theta
  - w^2 \frac23 X_0 (1-X_0^2)\sin\left( \frac\kappa2 \right)\cos\left( \theta \right)
  . \label{Eq:phi0(psi)}
\end{equation}
Finally, we substitute this into expressions \eqref{Eq:0:CPR} and \eqref{Eq:2:CPR}, expand up to $O(w^2)$ and after some simplifications obtain
\begin{eqnarray}
  \gamma(\theta) &=& \cos\left( \frac\kappa2 \right) \sin(\theta) - X_0 \sin\left( \frac\kappa2 \right) \cos(\theta) +
  \nonumber\\
  &+& Q \sin^2\left( \frac\kappa2 \right) \sin(2\theta)
  , \label{Eq:CPR(psi)}
\end{eqnarray}
where, for the sake of brevity, we have introduced
\begin{equation}
  Q \equiv \frac16 w^2 (1-X_0^2)^2
  . \label{Eq:Q.def}
\end{equation}

\section*{References are on a separate page}
\clearpage
\bibliography{SF,SFS,pi,QuComp}

\end{document}